\documentstyle[eqsecnum,aps,epsf]{revtex}
\begin{document}
\input epsf
\def\slash#1{#1\!\!\!/}
\draft
      
\title{Pion Cloud and Nucleon Mass in Finite Nuclei}
\author{ Andrew J. Harey\thanks{E-mail:harey@cmu.edu} and 
Leonard S. Kisslinger\thanks{E-mail:kissling@andrew.cmu.edu}}

\address{Department of Physics, Carnegie Mellon University, 
Pittsburgh, Pennsylvania 15213-3890}
\date{\today}
\maketitle

\begin{abstract}
We study the mass of a nucleon in a nucleus using a model recently introduced 
for a nucleon current with a pion cloud component. The method of QCD sum rules 
is used to determine the mass shifts for nucleons to first order in nuclear 
density in $\,^{16}{\text {O}}$ and $\,^{40}{\text {Ca}}$, with the pion 
propagator determined by an optical potential for pion-nucleus
scattering. Mass 
shifts due to the pionic contributions range from negligible in oxygen
to $-100 
\ {\text {MeV}}$ in calcium at central nuclear density, with larger shifts in 
the nuclear surface. The consistency of this model with chiral perturbation 
theory is demonstrated.
\end{abstract}


\newpage

\section{Introduction}

Until the pioneering work of Drukarev and Levin on the use of QCD sum rules in 
systems of finite nucleon density \cite{druklev}, descriptions of nucleons in 
nuclear matter had mostly been based on models of $N$-$N$ interactions 
\cite{bethe} or the direct use of meson exchange potentials in quantum 
hadrodynamics \cite{qhd}. These models have serious problems at both the 
short-distance and long-distance scales. At short distance one must treat the 
complex structure of hadrons, for which we now believe that QCD gives the 
correct description. At long-distance scales it seems that meson cloud effects 
must be included, and neither quantum hadrodynamics nor current treatments of 
QCD are adequate; while chiral perturbation theory \cite{lp,gl} is very 
successful in a quantitative treatment of pionic effects at low energies. 
Recently a model was developed for use in QCD sum rules with a nucleon current 
that has a Goldstone boson component, which has the main features of both 
microscopic QCD and chiral perturbation theory \cite{kissl1}. In the present 
paper we extend this model for the study of nucleons in finite nuclei.

The strongest experimental evidence for the utility of introducing an explicit 
meson cloud component for the nucleon is the data on sea quark distributions 
obtained from Drell-Yan experiments. The NMC/CERN experiments \cite{nmc} have 
given evidence for violation of the Gottfried sum rule
\begin{equation} \label{gott}
\int_{0}^{1} {dx \over x}\left[F_{2}^{p}\left(x\right) -
               F_{2}^{n}\left(x\right)\right]
 = {1 \over 3} \int_{0}^{1}\left\{\left[u_{p}^{v}\left(x\right) 
               - d_{p}^{v}\left(x\right) \right] 
               + 2\left[\bar{u}_{p}^{v}\left(x\right) 
               - \bar{d}_{p}^{v}\left(x\right)\right] \right\}dx.
\end{equation}
If the up-type and down-type sea quark distributions are equal, as would be 
expected from QCD, then the value of this integral is $1/3$. The NMC result is 
$0.24 \pm 0.016$ at $Q^{2}=5 \ {\text {GeV}}^{2}$ over the interval
$0.004 \leq 
x \leq 0.8$. Similar effects were found in Drell-Yan measurements from 
Fermilab/E866 \cite{e866}, where the ratio of down to up sea quark 
distributions is much larger than 1.0. Both of these results differ from the 
predictions of perturbative \cite{pqcd} and nonperturbative \cite{nqcd} QCD 
calculations. One approach to account for this discrepancy in sea quark 
distributions in hadronic models is to include a meson cloud for the nucleon 
\cite{sul,st,mmp}. Our approach is to treat the pion as a basic
Goldstone boson 
field at low energy, along with quarks and gluons, with separation of 
long-distance from medium- and short-distance effects.

Another strong motivation for introducing a correlator with a Goldstone boson 
component for use in QCD sum rules is given by chiral perturbaton theory. It 
was shown almost two decades ago \cite{gz} that the leading nonanalytic 
contribution (LNAC) to the nucleon mass in chiral perturbation theory is a 
m$_\pi^3$ correction, with the lowest-order chiral logs not contributing. In 
the standard formulation of QCD sum rules one does not obtain such a m$_\pi^3$ 
term \cite{lccg}. In Ref \cite{kissl1} it was shown not only that one obtains 
the m$_\pi^3$ LNAC, but that the known value of the mass shift (15 MeV) can be 
used to help determine the parameters of the model, in spite of the fact that 
the sum rule method is only accurate to about 10\% for the nucleon mass. It is 
not possible to obtain the LNAC  m$_\pi^3$ term with the usual quark fields 
correlator \cite{lccg,kissl1}. We make use of this result in the present work.

It should also be noted that in the QCD sum rule method the microscopic QCD 
calculation gives $m_{\pi}^{2}\ln(m_{\pi}^{2})$ terms for the nucleon in
vacuum 
and also $\rho m_{\pi}$ terms for nucleons in the nuclear medium, both in 
contradiction to chiral perturbation theory. It was shown that both for the 
vacuum \cite{lccg} and for nuclear matter \cite{birse} a careful treatment of 
the contunuum part of the phenomenological correlator enables one to cancel 
these terms, which are inconsistent with chiral perturbation theory.  We made 
use of this cancellation of unwanted logs in the formulation of our model 
\cite{kissl1} and use this mechanism in the present work. 
 
We review Goldstone boson/QCD model of Ref. \cite{kissl1} for a nucleon in the 
vacuum in Section 2. In Section 3 the surface effects on the pion propagator 
are quantified using a p-wave $\pi A$ optical potential where the 
$\Delta(1232)$ resonance is dominant. We extend the model for the microscopic 
correlator from zero to finite density in Section 4, with the goal of 
calculating nucleon mass shifts due to the pion cloud effects in small 
symmetric finite nuclei. The sum rules are then constructed and evaluated in 
the final sections.

\section{Review of Nucleon Correlator with Pion Cloud in Vacuum}

In this section we review the model for the nucleon correlator with explicit 
Goldstone boson fields developed in Ref \cite{kissl1}. The nucleon current has 
the form \begin{equation}
\eta^{N}(x)=c_{1}\eta^{N,0}(x)+c_{2}\eta^{N,\pi}(x)
\end{equation}
where the constants $c_{i}$ are the amplitudes of the composite field operator 
without and with the Goldstone boson field. This operator is used to construct 
the nucleon correlator. We ignore possible contributions to the nucleon from 
the strange part of the meson cloud. The proton field operator without
the pion 
cloud is chosen to be the current \cite{ioffe} 
\begin{equation} \label{eq:ioffec}
\eta^{p,0}(x)=\epsilon^{abc}\left[u^{a}(x)^{T}C\gamma_{\mu}u^{b}(x)\right]
               \gamma_{5} \gamma^{\mu}d_{c}(x),
\end{equation}
where $a,b,c$ label the color indices, and $u(x), d(x)$ are the up and down 
quark fields. $C$ is the charge conjugation operator. The lowest-energy 
contribution to the phenomenological dispersion relation for the correlator, 
the pole term giving the proton intermediate state, depends on the transition 
matrix element 
\begin{equation}
\left<0\left|\eta^{p,0}(p)\right|{\text {proton}}\right>=\lambda_{p}v(p),
\end{equation}
where $\lambda_{p}$ is a structure constant and the Dirac spinor $v(p)$ is 
normalized by
\begin{equation}
\bar{v}(p)v(p)=2m.
\end{equation} 
The neutron current $\eta^{n,0}(x)$ can be obtained by the interchange of up 
and down quark fields in Equation (\ref{eq:ioffec}). The nucleon current 
including the explicit Goldstone boson field is taken to be 
\begin{eqnarray}\label{eq:picurr}
\eta^{p,\pi}(x)&=&{1 \over \lambda_{\pi}^{2}} 
                   \partial_{\alpha}\phi_{\pi}(x) \cdot \tau 
                   \gamma^{\alpha}\gamma^{\beta} \eta^{N,0}(x),
\end{eqnarray}
where $\phi_{\pi}(x)$ is a massless pion field, $\tau$ is the I-spin operator 
and $\lambda_{\pi}$ is a $D=1$ scale parameter. The structure constant of this 
current is defined by the relation
\begin{equation}
\left<0\left|\eta^{p,\pi}(x)\right|{\text {proton}}\right>=
              \lambda_{p}'v(p).
\end{equation} 
The p-wave coupling of the pion to the nucleon current and experience with 
hybrid mesons \cite{kissl2} suggests that 
$\lambda_{p}'^{2} \ll \lambda_{p}^{2}$, so we expect little contribution to 
the phenomenological expression for our nucleon correlator from the resulting 
pole due to its very small residue. 

The full correlator for the proton with the pion cloud is
\begin{eqnarray} 
\label{eq:vacpcor}
\Pi^{p}(p)&=&i\!\int \!d^{4}\!x \,{\text e}^{i p \cdot x}\left<0\left| 
        T\left[\eta^{p}(x)\bar{\eta}^{p}(0)\right]\right|0\right> \nonumber \\
&=&c_{1}^{2}i \!\int \!d^{4}\!x \,{\text e}^{i p \cdot x}
\left<0\left| T\left[\eta^{p,0}(x)\bar{\eta}^{p,0}(0)\right]\right|0\right> +
(1-c_{1}^{2}) i \! \int \!d^{4}\!x \,{\text e}^{i p \cdot x}
\left<0\left| T\left[\eta^{p,\pi}(x)\bar{\eta}^{p,\pi}(0)\right]\right|0\right>
\nonumber \\
&=&c_{1}^{2}\Pi^{(p,0)}(p)+(1-c_{1}^{2})\Pi^{(p,\pi)}(p).
\end{eqnarray} 
With the appropriate changes of current and isospin considerations, the
neutron 
correlator is obtained. This model focuses on the long-range effects of the 
pion cloud and therefore does not include pion-quark interactions and the 
coupling between currents that include the pion cloud and currents that do 
not. It is assumed that these short-range contributions are already accounted 
for in QCD through the condensates, and their inclusion in the model would 
result in double counting some of the pionic effects on the nucleon correlator.

The part of the proton correlator without the pion cloud can be found in 
Ref.\cite{ioffe}. The nucleon correlator with the meson cloud for the 
microscopic evaluation of the sum rule has the form 
\begin{equation}
\label{picor}
\Pi^{(p,\pi)}(x)=D_{f,\alpha \beta}^{\pi}(x)\gamma^{\alpha}
\Pi^{(p,0)}(x) \gamma^{\beta},
\end{equation} 
where $\Pi^{(p,0)}(x)$ is the coordinate space nucleon correlator without the 
pion cloud \cite{ioffe} and in the chiral limit $D_{f,\alpha \beta}^{\pi}(x)$ 
has the form
\begin{equation} \label{eq:dfab}
D_{f,\alpha \beta}^{\pi}(x)={1 \over \pi^{2} x^{4}}\left( {4 x_{\alpha}
x_{\beta} \over x^{2}} - \delta_{\alpha \beta} \right).
\end{equation}

In momentum space the nucleon correlator with the pion cloud is 
\begin{mathletters}
\begin{eqnarray}
\Pi^{(p,\pi)}(p)&=&-2 \! \int \!{d^{4}\!k \over \left(2 \pi\right)^{4}}
{\left(\slash{p} - \slash{k}\right) \Pi^{(p,0)}(k) \left(\slash{p} -
\slash{k}\right) \over \left(p - k\right)^{2}} \\
&=&\slash{p}\Pi^{(p,\pi)}_{{\text {odd}}}(p) + 
            \Pi^{(p,\pi)}_{{\text {even}}}(p).
\end{eqnarray}
\end{mathletters}

\begin{figure} 
\begin{center}
\epsfxsize=250pt \epsfbox{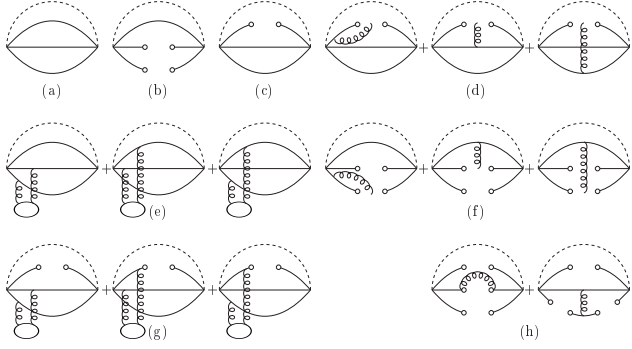}
\end{center}
\caption{The diagrams corresponding to the pion cloud contributions to
the nucleon correlator in vacuum.  The solid lines correspond to quarks,
the helical lines to gluons and the dashed lines to Goldstone bosons.
The circles and ovals correspond to the vacuum condensates.}
\label{fig:vacpidia}
\end{figure}

The pion cloud contributions to the nucleon correlator are shown in the 
diagrams of  Figure \ref{fig:vacpidia}, and in momentum space are
\begin{mathletters}
\begin{eqnarray}
\Pi_{1a}(p)&=&{i \over 2^{12} 3 \cdot 5
\pi^{6}\lambda_{\pi}^{4}}p^{8}\ln(-p^{2})\slash{p} \\
\Pi_{1b}(p)&=&{i \left<\bar{q}q\right>^{2} \over 2^{3} 3 \pi^{2}
\lambda_{\pi}^{4}} p^{2}\ln(-p^{2})\slash{p} \\ 
\Pi_{1d}(p)&=&{i \left< g_{s}^{2}G \cdot G \right> \over 2^{11} 3
\pi^{6} \lambda_{\pi}^{4}} p^{4}\ln(-p^{2})\slash{p} \\
\Pi_{1e}(p)&=&{-i \left<\bar{q}q\right> \left< \bar{q} g_{s}^{2}\sigma
\cdot G q \right> \over 2^{5} \pi^{2} \lambda_{\pi}^{4}} \ln(-p^{2})
\slash{p}.
\end{eqnarray}
\end{mathletters} 
Details are given in Ref. \cite{kissl1}. The usual  QCD sum rule methods are 
used, as in Ref. \cite{ioffe}, but with a modification of the treatment of the 
continuum given by the work of Ref. \cite{lccg}. In addition to the
calculation 
of the nucleon mass the magnetic dipole moments of the nucleons were 
calculated. From the comparison of the results for the magnetic dipole moment 
to the result without the pion cloud component \cite{is} it was estimated 
that the pion cloud component has roughly the same probability as the 
cloudless component, or $c_1^2 \simeq c_2^2 \simeq 0.5$.

\begin{figure} 
\begin{center}
\epsfxsize=125pt \epsfbox{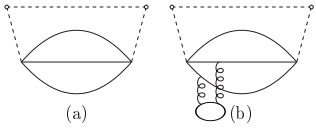}
\end{center}
\caption{The diagrams corresponding to the lowest-dimension chiral
corrections to the nucleon correlator in vacuum}
\label{fig:chivac}
\end{figure}

An important observation is that the diagrams shown in Figure \ref{fig:chivac}
 gives the main LNAC $m_\pi^3$ terms. The largest contribution is from the 
 first term shown in Figure \ref{fig:chivac}a, which was shown to give the 
 following contribution to the nucleon mass 
\begin{eqnarray}
\label{cp16}
   \Delta M_p & = & -\frac{e m_\pi^3 g_A^2}{160 \beta^2 f_\pi^2} \nonumber \\ 
              & \simeq & -20 \ \text{MeV},
\end{eqnarray}
with the choice of $\lambda_\pi^2$ = M$_p^2$/g$_{\pi N}$, which is quite 
reasonable. 

The contribution to the nucleon mass from the pion cloud was found to be 
negligible, consistent with the assumption 
$\lambda_{p}'^{2} \ll \lambda_{p}^{2}$. We use this value for  
$\lambda_\pi^2$ in the present work. This completes the model within the 
uncertainty of the probability of the meson cloud component.

\section{Pion Propagator in Finite Nuclei}

Extending the model for the nucleon correlator with the pion cloud 
from vacuum to finite nuclei requires an appropriate modification 
of the pseudo-Goldstone boson field in medium. In order to be able 
to able to apply the model at high momentum transfer, which is needed 
for various applications, we make use of conventional multiple scattering 
formalism [see, e.g., Ref. \cite{eiskol} for a review] rather than chiral 
perturbation theory. This is a mean field method in which the pion 
propagator in the nucleon medium is given by the pion optical potential.

The propagator for a free massless pion has the form 
\begin{equation} \label{eq:freeprop}
D^{\pi}_{f}(k)={1 \over k^2 + i\epsilon}.
\end{equation}
In the medium the pion acquires an effective mass given by the optical 
potential. The propagator for a pion with energy $k_{0}$ in a nucleus of 
mass number $A$ is 
\begin{equation} \label{eq:medprop}
D^{\pi}_{A}(k)={1 \over k^{2}+\Pi(k_{0},\vec{k};\rho,A)+i\epsilon}
\end{equation}
where $\Pi(k_{0},\vec{k};\rho,A)$ is the pion self-energy in the medium 
and $\rho$ is the baryon density of the matter. The pion self energy is 
related to the pion-nucleus optical potential $V_{\text {opt}}(k_{0})$ by 
\begin{equation}
\Pi(k_{0},\vec{k},\vec{k}')=
-2k_{0}\left<\vec{k}'\left|V^{\pi A}_{\text {opt}}(k_{0})\right|\vec{k}\right>,
\end{equation} 
where $\vec{k}$ and $\vec{k}'$ are respectively the incoming and outgoing 
three-momenta of the pion. We begin constructing the $\pi A$ optical potential 
by looking at the partial wave expansion of the $\pi N$ scattering amplitude 
\begin{equation}
{\mathcal F} \left(\vec{k}',\vec{k}\right)=
\sum_{I}\hat{P}_{I}\left(\sum_{l}\left[\left( l+1\right)f_{I,l^{+}}(\omega)
+lf_{I,l^{-}}(\omega)\right]P_{l}(\cos \theta) 
- {\text i}\vec{\sigma}\cdot \left(\hat{k}'\cdot\hat{k}\right)\sum_{l} 
\left[f_{I,l^{+}}(\omega)-f_{I,l^{-}}(\omega)\right]P_{l}'(\cos \theta)\right).
\end{equation} 
The partial wave amplitude is related to the S-matrix by
\begin{equation}
f_{\alpha}={1 \over 2{\text i}|\vec{k}|}\left(S_{\alpha}(\omega)-1\right)
\end{equation}
where the scattering matrix is found using the experimentally obtained phase 
shifts $\delta_{\alpha}(\omega)$ as 
\begin{equation}
S_{\alpha}(\omega)={\text e}^{2{\text i}\delta_{\alpha}(\omega)}.
\end{equation} 
In this notation $\alpha$ indexes the isospin $I$, orbital angular
momentum $l$ 
and total angular momentum $J$ of the partial wave. The phase shifts are 
expected to depend on momentum proportional to $k^{2l+1}$, and we define 
the constants $a_{\alpha}$ as 
\begin{equation}
a_{\alpha}={\delta_{\alpha} \over k^{2l+1}}.
\end{equation} These constants are the so-called scattering lengths 
($a_{2I}$) for s-waves and the scattering volumes ($b_{2I,2J}$) for 
p-waves. These  parameters for d-wave and higher order partial waves 
are very small and have negligible effect on $\pi N$ physics, so we 
truncate the expansion of the scattering amplitude at $l=1$. Then 
the $\pi N$ scattering amplitude can take the form 
\begin{equation} \label{eq:pinf}
{\mathcal F}(\vec{k}', \vec{k})=a(k_{0}) + 
a_{I}(k_{0})\left(\vec{t}\cdot\vec{\tau}\right) 
+ \left[b(k_{0}) 
+ b_{I}(k_{0})\left(\vec{t}\cdot\vec{\tau}\right)\right]\vec{k}'\cdot\vec{k} 
+ {\text i} \left[b_{sf}(k_{0})
+ b_{Isf}(k_{0})\left(\vec{t}\cdot\vec{\tau}\right)\right] 
\vec{\sigma}\cdot\left(\vec{k}'\times \vec{k}\right),
\end{equation} 
where $\vec{t}$ and $\vec{\tau}$ are the isospin matrices for the pion and 
the nucleon. The energy-dependent coefficients can be related to the
scattering 
volumes and lengths at threshold; for example, 
$b={1\over 3}\left(4b_{33}+2b_{13}+ 2b_{31}+b_{11}\right)$. Because we are 
concerned in this work with spin/isospin symmetric nuclei, the isospin 
dependent and spin-flip parts of the scattering amplitude vanish. 
Furthermore, the p-wave contibution is considerably larger than the s-wave 
contribution at low and medium energies because of the dominance of the 
$\Delta(1232)$ $3,3$ resonance in $\pi N$ scattering. We are then left 
with the simple scattering amplitude for the spin and isospin averaged 
scattering of a pion off of a nucleon, 
\begin{equation}
\label{eq:f}
{\mathcal F}(\vec{k}', \vec{k})=b(k_{0})\vec{k}'\cdot\vec{k}.
\end{equation} 

In order to take into consideration the off-energy-shell effects in our 
optical potential, we modify the result of (~\ref{eq:f}) by considering 
the p-wave t-matrix for the $\pi N$ scattering in the separable form 
\begin{equation}
\label{tpin}
\left<\vec{k}'\left|t_{\pi N}^{p-{\text{wave}}}
\left(k_{0}\right)\right|\vec{k}\right>
= -b(k_{0})g(\vec{k}')g(\vec{k})\vec{k}'\cdot\vec{k},
\end{equation} 
where $g(\vec{k})$ is a form-factor that roughly parametrizes the 
inelastic channels in the $\pi N$ scattering at high energies. 
$g(\vec{k})$ is chosen be unity when $\bar{k}$ is at the on-energy-shell 
value, and to approach zero as $\bar{k}$ gets large, where 
$\bar{k} \equiv |\vec{k}|$. We use a monopole parametrization of the 
form factor 
\begin{equation}
g(\vec{k})={m_{\pi}^{2}+\Lambda^{2} \over |\vec{k}|^{2}+\Lambda^{2}}, 
\end{equation} 
with vertex factor $\Lambda$ taken to be 700 MeV, for the description of the 
off-shell properties of the amplitude.

The energy dependence of the function $b(k_{0})$ is modelled by using the 
Breit-Wigner form for the $\Delta(1232)$, giving 
\begin{equation}
b(k_{0}, \bar{k})=\left({f^{2}\omega_{\Delta} \over 3 \pi
m_{\pi}^{3}}\right){k_{0}^{-1} \over \omega_{\Delta}-k_{0}-{i \over
2}\Gamma_{\Delta}\left(\bar{k},k_{0}\right)},
\end{equation}
where $\omega_{\Delta}$ is the energy of the $3,3$ resonance and $f$ is the 
pseudovector $\pi N$ coupling constant. We model the resonance so that the 
three-momentum dependence of the width $\Gamma_{\Delta}$ is cut off at low 
momenta and at the momentum where the resonance is saturated: 
\begin{equation}
\Gamma_{\Delta}\left(\bar{k},k_{0}\right)
={8 \over 3}{f^{2} \over 4 \pi}{\omega_{\Delta} \over m_{\pi}^{2}}
{1\over k_{0}}\left\{ 
\begin{array}{ll}
\bar{k}_{m}^{3}&\bar{k}\leq \bar{k}_{m}\\ \\
\bar{k}^{3}&\bar{k}_{m}\leq\bar{k}\leq\bar{k}_{\Delta} \\ \\
\bar{k}_{\Delta}^{3}&\bar{k}\geq\bar{k}_{\Delta} 
\end{array}\right. .
\end{equation} 
The lower limit starting at $\bar{k}_{m}=30$ MeV ensures that the 
$\Delta(1232)$ does not vanish even at very low pion momenta, and the 
upper limit at the three-momentum $\vec{k}_{\Delta}$ where the Breit-Wigner 
curve peaks provides a mechanism for the resonance to disappear at high 
energies as expected.

To obtain the pion-nucleus optical potential for elastic scattering we make 
use of three approximations: 
\begin{itemize}
\item low density
\item impulse
\item local density.
\end{itemize}
The low density and impulse approximations lead to the standard
first-order in $\rho$ form for the optical potential, and the local
density approximation is used for finite nuclei to treat the interaction
at the nuclear surface as a function of $\rho$. The p-wave pion optical
potential for elastic scattering by an $A$-nucleon nucleus can then be
written to first order in density as
\begin{equation}
\left<\vec{k}'\left|V^{\pi A}_{\text{opt}}\right|\vec{k}\right>= A
\left<\vec{k}'\left|t_{\pi A}^{p{\text {-wave}}}\right|\vec{k}\right>
\tilde{\rho}\left(\vec{k}-\vec{k}'\right).
\end{equation} 
$\tilde{\rho}(\vec{k}-\vec{k}')$ is the Fourier transform of the ground
state nuclear density $\rho$(r). The resulting optical potential in
coordinate space, using Eq.(\ref{tpin}), has the  form
\begin{equation}
\label{vopt}
             V(\vec{r}) = A c(E_\pi) \bigtriangledown \cdot \rho(r)
\bigtriangledown.
\end{equation}
This is the so-called Kisslinger potential \cite{lsk} without the s-wave
scattering term. It has been shown to give an accurate treatment of
medium-energy pion-nucleus elastic scattring \cite{eiskol}. Due to the
p-wave nature of the interaction derived from the $\Delta(1232)$
resonance the interaction takes place mainly on the surface, and indeed
vanishes in the interior of the nucleus if a constant density is used.
This will be seen to be an important aspect of our present work.

We model the spatial dependence of the nuclear density as a second order
harmonic oscillator
\begin{equation} \label{eq:density}
\rho (r)=\rho_{0}\left(1+\alpha\left({r \over c}\right)^{2}\right)
\exp\left(-\left({r \over c}\right)^{2}\right),
\end{equation}
where $\rho_{0}$ is the density of nuclear matter, $c$ is a size
parameter, and $\alpha$ provides some control over the shape, allowing
for modelling of heavier nuclei (like $\,^{40}{\text {Ca}}$) and more
complicated distributions (for example, $\,^{16}{\text {O}}$ with a
local minimum at the center).

We now turn our attention to the pion propagator in medium. The
propagator (\ref{eq:medprop}) is expanded in density as
\begin{equation}
D^{\pi}_{A}(k)=\left[D^{\pi}_{A}(k)\right]_{\rho_{0}=0} + \sum_{\vec{q}}
\tilde{\rho}(\vec{q}) \left[{\delta \over \delta
\tilde{\rho}(\vec{q})}D^{\pi}_{A}(k)\right]_{\rho_{0}=0} 
 + {1 \over 2!}\sum_{\vec{q}_{1}}
               \sum_{\vec{q}_{2}}\tilde{\rho}(\vec{q}_{1})
\tilde{\rho}(\vec{q}_{2}) \left[{\delta \over \delta
\tilde{\rho}(\vec{q}_{1})}{\delta \over \delta
\tilde{\rho}(\vec{q}_{2})}D^{\pi}_{A}(k)\right]_{\rho_{0}=0} + \cdots
\end{equation} 
where the first term is just the free pion propagator
(\ref{eq:freeprop}) and $\vec{q}$ is the momentum transfer in the
scattering. The scattering angles are averaged over, and the pion
propagator in the nucleus takes the form
\begin{equation} \label{eq:medpiprop}
D^{\pi}_{A}(k)=D^{\pi}_{f}(k) + {8 \pi^{3/2} (A-1) \rho_{0} f^{2}
\omega_{\Delta} \over m_{\pi}^{2} c} g\left(\bar{k}\right)^{2}
F_{1}\left(\bar{k},\alpha, c \right) \left(\omega_{\Delta}-k_{0}-{i
\over 2}\Gamma_{\Delta}\left(\bar{k}, k_{0}\right)\right)^{-1}
D^{\pi}_{f}(k)^{2} + {\mathcal O}\left(\rho_{0}^{2}\right),
\end{equation}
with the nuclear structure information contained in the function
\begin{eqnarray}
F_{1}\left(\bar{k},\alpha, c \right) & = & \left(1 + {3 \alpha \over
2}\right) \left[\left( 1 - {2 \over c^{2}\bar{k}^{2}}\right) + \left(2
c^{2} \bar{k}^{2} + 3 + {2 \over c^{2}\bar{k}^{2}}
\right)\exp\left(-c^{2}\bar{k}^{2}\right)\right] \nonumber \\
& & \mbox{ } - \alpha \left[ \left(1 - {3 \over c^{2}\bar{k}^{2}}\right)
+ \left(2 c^{4} \bar{k}^{4} + 4 c^{2}\bar{k}^{2} + 5 + {3 \over
c^{2}\bar{k}^{2}}\right) \exp\left(-c^{2}\bar{k}^{2}\right)\right].
\end{eqnarray}

\section{The Nucleon Correlator with Pion Cloud in Medium}

To construct the in-medium nucleon correlator with the meson field, we
start by rewriting the vacuum matrix element of the product
$\eta(x)\bar{\eta}(x)$ in Eq.(~\ref{eq:vacpcor}) as a nuclear matrix
element so that the proton current correlator takes the form
\begin{eqnarray} 
\label{eq:medpcor}
\Pi_{A}^{p}(p)&=&i\!\int \!d^{4}\!x \,{\text e}^{i p \cdot
x}\left<A\left|
T\left[\eta^{p}(x)\bar{\eta}^{p}(0)\right]\right|A\right> \nonumber \\
&=&c_{1}^{2}i \!\int \!d^{4}\!x \,{\text e}^{i p \cdot x}
\left<A\left|
T\left[\eta^{p,0}(x)\bar{\eta}^{p,0}(0)\right]\right|A\right> + c_2^{2}
i \! \int \!d^{4}\!x \,{\text e}^{i p \cdot x} \left<A\left|
T\left[\eta^{p,\pi}(x)\bar{\eta}^{p,\pi}(0)\right]\right|A\right>
\nonumber \\
&=&c_{1}^{2}\Pi_{A}^{(p,0)}(p)+c_{2}^{2}\Pi_{A}^{(p,\pi)}(p),
\end{eqnarray} 
where $\left|A\right>$ is the ground state of the finite nucleus. The
part of the in-medium proton correlator with the pseudo-Goldstone field
is
\begin{eqnarray} 
\label{eq:medxcor}
\Pi_{A}^{(p,\pi)}(x)&=&\epsilon^{abc}\epsilon^{a'b'c'}{1 \over
\lambda_{\pi}^{4}}\left\{{2 \over 3} \left(\partial_{\alpha}
\partial_{\beta}D_{A}^{\pi^{0}}(x)\right) \gamma_{\alpha}
\gamma^{\mu}S_{d,A}^{cc'}(x)\gamma^{\nu} \gamma_{\beta} {\text
{Tr}}\left[ S_{u,A}^{bb'}(x)\gamma_{\nu}C S_{u,A}^{aa'T}(x) C
\gamma_{\mu}\right] \right.
\nonumber \\
& & \left. \mbox{}+{4 \over 3} \left(\partial_{\alpha}
\partial_{\beta}D_{A}^{\pi^{+}}(x)\right) \gamma_{\alpha}
\gamma^{\mu}S_{u,A}^{cc'}(x)\gamma^{\nu} \gamma_{\beta} {\text
{Tr}}\left[ S_{d,A}^{bb'}(x)\gamma_{\nu}C S_{d,A}^{aa'T}(x) C
\gamma_{\mu}\right] \right\} \nonumber \\
& & \mbox{}+\left[{\text {4-quark \ terms}}\right] \nonumber \\
&=&{1 \over \lambda_{\pi}^{4}} \left\{ {2 \over 3}D_{A,\alpha
\beta}^{\pi^{0}}(x) \gamma^{\alpha} \Pi_{A}^{(p,0)}(x) \gamma^{\beta} +
{4 \over 3}D_{A,\alpha \beta}^{\pi^{+}}(x) \gamma^{\alpha}
\Pi_{A}^{(n,0)}(x) \gamma^{\beta} \right\}.
\end{eqnarray}
This is simply the expression for the meson cloud part of the
correlator, given in Ref. \cite{kissl1}, where the quark propagator and
pion propagator in vacuum, $S_{q}^{ab}(x)$ and $D_f^{\pi}(x)$, are
replaced by  $S_{q,A}^{ab}(x)$ and $D_{A}^{\pi}(x)$, the quark and pion
propagators in the nucleus, respectively. For example,
\begin{eqnarray}
\label{qprop}
    S_{q,A}^{ab}(x) & = &  
      \left<A\left| T\left[q^a(x)\bar{q}^{b}(0)\right]\right|A\right>. 
\end{eqnarray}
$D_{A,\alpha \beta}^{\pi}(x)$ is the nuclear analogue to
Eq.(\ref{eq:dfab}), replacing the free pion propagator $D_{f}^{\pi}(x)$
with $D_{A}^{\pi}(x)$. As with the case in vacuum, the neutron
correlator $\Pi_{A}^{(n,0)}(x)$ is found simply by an interchange of up
and down quark fields in the current $\eta(x)$.  With a spin/isospin
symmetric nucleus in the chiral limit, the in-medium correlator with a
pion cloud has the form analogous to Eq.(\ref{picor}),
\begin{equation} \label{eq:medmast}
\Pi_{A}^{(p,\pi)}(x)={2 \over \lambda_{\pi}^{4}} D_{A,\alpha
\beta}^{\pi}(x) \gamma^{\alpha} \Pi_{A}^{(p,0)}(x) \gamma^{\beta}.
\end{equation} 

For simplicity we work in momentum space, where the  in-medium proton
correlator has the form
\begin{equation} \label{eq:medcormom}
\Pi_{A}^{p}(p)=c_{1}^{2}\Pi_{A}^{(p,0)}(p)
   -\left( 1-c_{1}^{2} \right){2 \over \lambda_{\pi}^{4}} 
    \int{d^{4}\!k \over \left(2 \pi\right)^{4}} 
     D_{A}^{\pi}(k) \slash{k} \Pi_{A}^{(p,0)}(p-k) \slash{k} 
\end{equation}
where we made use of the Fourier transform of $D_{A,\alpha
\beta}^{\pi}(x):$
\begin{equation}
D_{A,\alpha \beta}^{\pi}(p)=-p_{\alpha}p_{\beta}D_{A}^{\pi}(p).
\end{equation}

Since the nucleus has a four-momentum $u^{\mu}$, breaking Lorentz
invariance, there are additional Dirac structures in the nucleon
correlator. Under the assumption that the nuclear ground state is
invariant under parity and time-reversal, it was shown \cite{druklev}
that the in-medium nucleon correlator is
\begin{equation}
\Pi_{A}^{p}(p)=\Pi_{A}^{(s)}(p^2,p\cdot u) 
               + \Pi_{A}^{(p)}(p^2,p\cdot u) \slash{p} 
               + \Pi_{A}^{(u)}(p^2,p\cdot u) \slash{u}.
\end{equation} 
The individual scalar functions $\Pi_{A}^{(n)}(p^2,p\cdot u)$ depend
only on the Lorentz invariants $p^{2}$ and $p \cdot u$ and can be
projected out of the full nucleon correlator using the following
formulae \cite{druklev,cofugr}: \begin{eqnarray}
\Pi_{A}^{(s)}(p^2,p\cdot u)&=&{1 \over 4} 
  {\text {Tr}}\left[ \Pi_{A}^{p}(p) \right] \nonumber \\
\Pi_{A}^{(p)}(p^2,p\cdot u)&=&{1 \over p^{2} - \left( p \cdot u
\right)^{2}} \left[ {1 \over 4} {\text {Tr}}\left[ \slash{p}
\Pi_{A}^{p}(p) \right] - {p \cdot u \over 4} {\text {Tr}}\left[
\slash{u} \Pi_{A}^{p}(p) \right] \right] \\
\Pi_{A}^{(u)}(p^2,p\cdot u)&=&{1 \over p^{2} - \left( p \cdot u
\right)^{2}} \left[ {1 \over 4} {\text {Tr}}\left[ \slash{u}
\Pi_{A}^{p}(p) \right] - {p \cdot u \over 4} {\text {Tr}}\left[
\slash{p} \Pi_{A}^{p}(p) \right] \right]. \nonumber 
\end{eqnarray} 

We define in-medium matrix elements of an operator 
${\mathcal O}\left( x \right)$ as an expansion in local nuclear 
density $\rho$
\begin{eqnarray}
\label{op}
 \left< {\mathcal O}\left( x \right) \right>_{\rho}& = &
 \left< A \left| {\mathcal O}\left( x \right) \right| A \right> 
 - \left< 0 \left|{\mathcal O}\left( x \right) 
                    \right| 0 \right> \\ \nonumber 
      & = &  \rho \left< N \left| {\mathcal O}\left( x 
                    \right) \right| N \right> + \cdots ,
\end{eqnarray}
where $\left< N \left| {\mathcal O}\left( x \right) \right| N \right>$
is the single-nucleon matrix element of ${\mathcal O}\left( x \right)$.
The in-medium quark condensate is related to the $\pi$-$N$ sigma
commutator by \cite{druklev} 
\begin{equation} \label{eq:qcondm}
\left< \bar{q} q \right>_{\rho} = 
          \rho {\sigma_{\pi N} \over m_{u}+m_{d}}.
\end{equation} 
We take $\sigma_{\pi N} \simeq 47 \ {\text {MeV}}$ \cite{gasser2}. The
breaking of the Lorentz invariance of the medium by the introduction of
the nuclear four-momentum $u^{\mu}$ leads to condensates with additional
Dirac structures . The dimension three in-medium vector quark
condensate, with the appropriate color and flavor weighting, is
proportional to the nuclear density
\begin{equation}
\left< q^{\dagger} q \right>_{\rho} = {3 \over 2} \rho.
\end{equation} 
The in-medium gluon condensate is found via the trace anomaly
\cite{henpas,tran} to be 
\begin{equation}
\left< g_{s}^{2} G_{\mu \nu} G^{\mu \nu} \right>_{\rho} 
  = -\rho 2^{6}39 \pi^{2} \ {\text {MeV}}.
\end{equation}
The values of the five-dimensional mixed quark-gluon condensates in
medium are not well known, although significant cancellation between
diagrams containing these matrix elements occurs much like similar terms
in the free nucleon correlator in the chiral limit. Furthermore, it was
shown in \cite{cofugr} that the nucleon sum rules are insensitive to the
values of these condensates, so we ignore terms proportional to the
mixed condensates in our sum rule analysis.

The nucleon correlator is analyzed in the rest frame of the nucleus,
where the nucleus four-momentum is $u=\left( M,0,0,0\right)$ with the
mass of the nucleus $M \approx A m$  (neglecting the average binding
energy per nucleon in the nucleus as small compared to the nucleon mass
$m$). In this frame $p \cdot u = 2 M p_{0}$, where $p_{0}$ is the energy
of the interpolating field. We carry out Borel transforms with respect
to $-p^2$, using the relationship
\begin{equation} \label{eq:constraint}
\left( {M \over A} + p \right)^{2} = 4 m^{2}.
\end{equation} 
In the nuclear rest frame, $p_{0} = {1 \over 2m}\left(3m^{2} -p^{2}
\right)$, so when $-p^{2} \rightarrow \infty$,  $p \cdot u \rightarrow
\infty$ while the ratio of the two invariants, analogous to the Bjorken
scaling variable in DIS, remains fixed and the OPE is applicable.

\begin{figure} 
\begin{center}
\epsfxsize=250pt \epsfbox{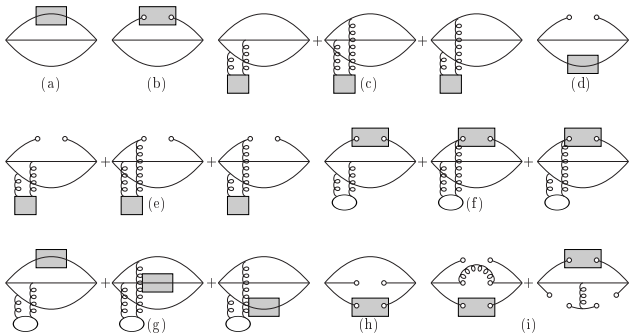}
\end{center}
\caption{The diagrams corresponding to the contributions to the nucleon
correlator in medium without the pion cloud. The shaded boxes represent
the interaction of the fields with the medium to first order in density.} 
\label{fig:mednpdia}
\end{figure}

\begin{figure} 
\begin{center}
\epsfxsize=250pt \epsfbox{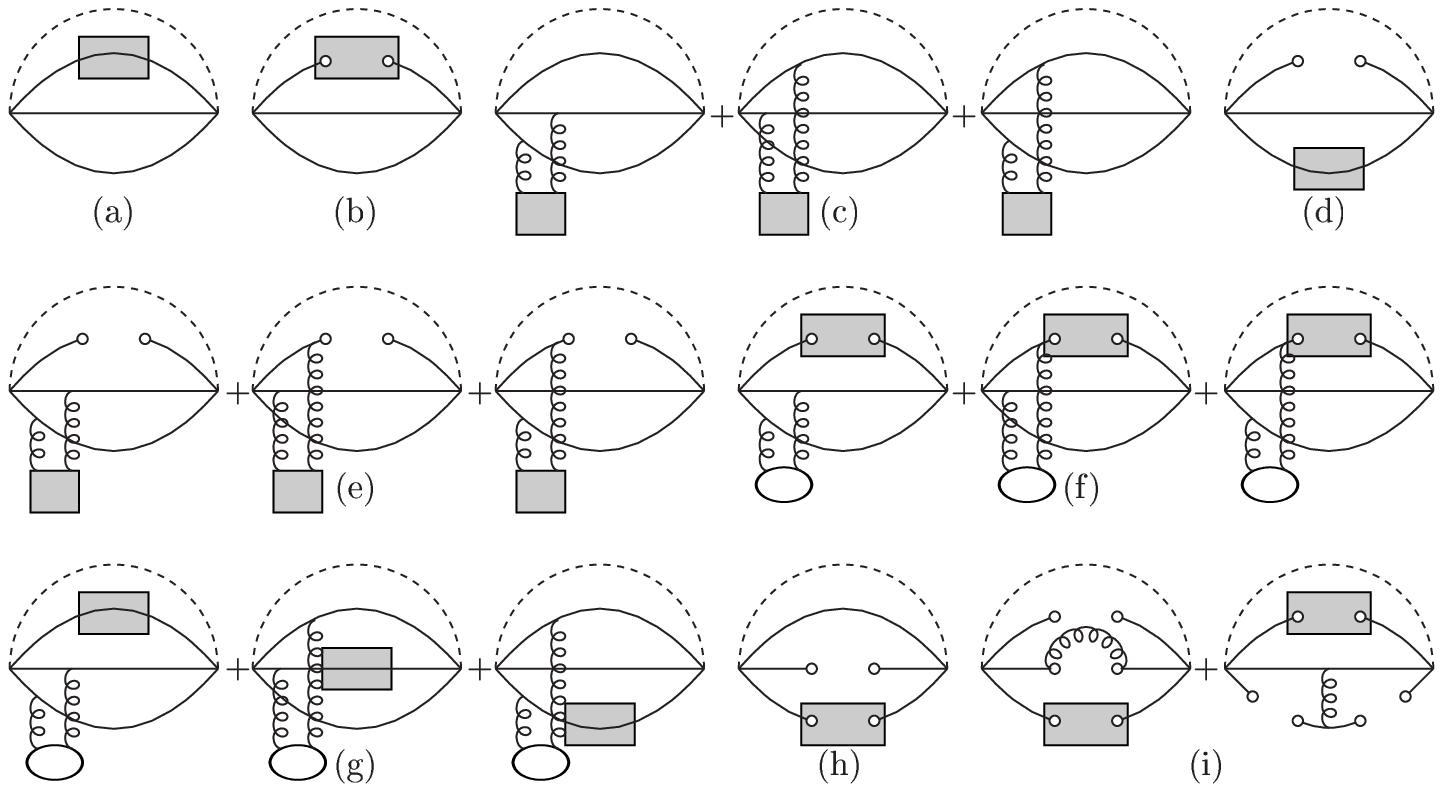}
\end{center}
\caption{The diagrams corresponding to the free pion cloud contributions
to the nucleon correlator in medium.} 
\label{fig:medpi1dia}
\end{figure}

\begin{figure} 
\begin{center}
\epsfxsize=250pt \epsfbox{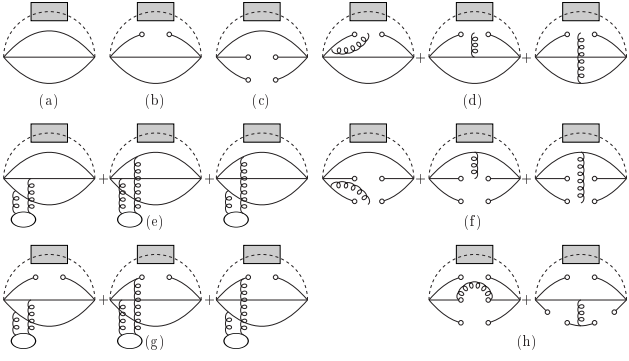}
\end{center}
\caption{The diagrams corresponding to the in-medium pion cloud
contributions to the nucleon correlator in medium.} 
\label{fig:medpi2dia}
\end{figure}

We restrict our calculation of the nucleon correlator to first order in
the nuclear density. We can then identify three distinct types of quark
current diagrams that contribute to the correlator: diagrams that
contribute to the part of the in-medium correlator without the
pseudo-Goldstone field (Figure \ref{fig:mednpdia}), diagrams that
include the pion field where the quarks and gluons interact with the
medium and the pion maintains its in-vacuum properties (Figure
\ref{fig:medpi1dia}), and diagrams where the quarks and glue interact
with the QCD vacuum and the pion scatters in the nucleus (Figure
\ref{fig:medpi2dia}). The shaded boxes in these diagrams represent
interaction of the field with the nuclear medium to first order in the
density expansion. Diagrams where the quarks and pions both interact
with the medium, or where the pion scattering includes two-body
correlations would come in at second order in density and are not
included in this work.

The first two sets of diagrams are most easily calculated in coordinate
space using (\ref{eq:medmast}), after which they are Fourier transformed
into momentum space. The contributions at first order in the density
from diagrams without the pion cloud are found to be 
\cite{druklev,cofugr}
\begin{eqnarray}
\Pi_{(2a,\rho)}(p)&=&{i \over 3 \pi^{2}} \left< q^{\dagger} q
\right>_{\rho} \left[ 2 \slash{u}p^{2}+  \slash{p} u\cdot p \right]
\ln\left(-p^{2}\right) \nonumber \\
\Pi_{(2b.\rho)}(p)&=&{i \over \left(2 \pi\right)^{2}} \left< \bar{q} q
\right>_{\rho} p^{2}\ln\left(-p^{2}\right) \nonumber \\
\Pi_{(2c,\rho)}(p)&=&-{i \over 2^{7} \pi^{4}} \left< g_{s}^{2}G\cdot G
\right>_{\rho} \slash{p}\ln\left(-p^{2}\right) \nonumber \\
\Pi_{(2d,\rho)}(p)&=&-{2^{2}i \over 3} \left< \bar{q} q \right> \left<
q^{\dagger} q \right>_{\rho} {u \cdot p \over p^{2}} \nonumber \\
\Pi_{(2e+2f,\rho)}(p)&=&-{i \over 3^{2}2 \pi^{2}} \left[ \left< \bar{q}
q \right> \left< g_{s}^{2} G \cdot G \right>_{\rho} + \left< \bar{q} q
\right>_{\rho} \left< g_{s}^{2} G \cdot G \right> \right] {1 \over
p^{2}} \nonumber \\
\Pi_{(2g,\rho)}(p)&=&{i \over 2^{5}3^{2}\pi^{2}} \left< q^{\dagger} q
\right>_{\rho} \left< g_{s}^{2} G\cdot G \right> \left[ 5 \slash{u}{1
\over p^{2}} -  2 \slash{p} {u\cdot p \over p^{4}}\right] \nonumber \\
\Pi_{(2h,\rho)}(p)&=&-{2^{2} i \over 3} \left< \bar{q} q \right> \left<
\bar{q} q \right>_{\rho} \slash{p} {1 \over p^{2}} \nonumber \\
\Pi_{(2i,\rho)}(p)&=&-{2^{2} 17 i \over 3^{3}} g_{s}^{2} \left< \bar{q} q
\right>^{2} \left< \bar{q} q \right>_{\rho} {1 \over p^{4}}.
\end{eqnarray} 
Here we have used a factorization approximation to evaluate the effects
of the four-quark and six-quark condensates, where to first order in
density we take $\left<\bar{q} \Gamma q  \bar{q} \Gamma q \right>_{\rho}
\approx \left< \bar{q} q \right> \left< \bar{q} q \right>_{\rho}$ and
$\left< \bar{q} \Gamma q \bar{q} \Gamma q \bar{q} \Gamma q
\right>_{\rho} \approx \left< \bar{q} q \right>^{2}\left< \bar{q} q
\right>_{\rho}$, respectively. While the factorization of these
condensates is unjustified in medium \cite{jk}, our primary concern is a
comparison of the nucleon mass shift due to the pion cloud terms and the
shift due to terms without the pion cloud, so we find the approximation
acceptable for the problem at hand.

The diagrams in Figure \ref{fig:medpi1dia}, where there are first-order
interactions of the quarks and gluons with the medium but the pions
interact only with the vacuum, provide the following contributions at
first order in the density:
\begin{eqnarray}
\Pi_{(3a,\rho)}(p)&=&-{i \over 2^{6} 3^{2} \pi^{4} \lambda_{\pi}^{4}}
\left< q^{\dagger} q \right>_{\rho} \left[ \slash{u}p^{2}+  3 \slash{p}
u\cdot p\right]p^{4}\ln\left(-p^{2}\right) \nonumber \\
\Pi_{(3c,\rho)}(p)&=&{i \over 2^{13} \pi^{6} \lambda_{\pi}^{4}} \left<
g_{s}^{2} G \cdot G \right>_{\rho} \slash{p} p^{4}\ln\left(-p^{2}\right)
\nonumber \\
\Pi_{(3g,\rho)}(p)&=&-{i \over 2^{8} 3^{3} \pi^{4} \lambda_{\pi}^{4}}
\left< q^{\dagger} q \right>_{\rho} \left< g_{s}^{2} G \cdot G \right>
\left[ 2 \cdot 5 \slash{u}p^{2}+  11 \slash{p} u\cdot
p\right]\ln\left(-p^{2}\right) \nonumber \\
\Pi_{(3h,\rho)}(p)&=&{i \over 2^{2} 3 \pi^{2} \lambda_{\pi}^{4}} \left<
\bar{q} q \right> \left< \bar{q} q \right>_{\rho} \slash{p}
p^{2}\ln\left(-p^{2}\right) .
\end{eqnarray} 
All diagrams which would be expected to contribute to
scalar structure of the nucleon correlator vanish.

The calculation of the terms represented in Figure \ref{fig:medpi2dia}
where the pions interact with the nucleus and the quarks and glue
interact within the QCD vacuum are more difficult to calculate. Instead
of calculating these diagrams in coordinate space as we did with the
other in-medium diagrams, which would require first a rather daunting
Fourier transform of the in-medium pion propagator (Eq.
\ref{eq:medpiprop}) and then an inverse transform of the resulting
expression, we perform the calculations in momentum space using Equation
(\ref{eq:medcormom}) as the starting point. For example, the term
corresponding to Figure \ref{fig:medpi2dia}a gives the linear-density
contribution
\begin{eqnarray*}
\Pi_{(4a,\rho)}(p)&=&{\zeta i \over 2^{10} \pi^{8} \lambda_{\pi}^{4}}
\int \! d^{3}\vec{k} \left(g(\bar{k})\right)^{2}
F_{1}\left(\bar{k},\alpha,c\right) \\
 & & \times \int \! dk_{0} k_{0} {\left(\omega_{\Delta} -
k_{0}\right)k_{0} + {i \over 2}\Gamma_{\Delta}\left(\bar{k}\right) \over
\left(\omega_{\Delta} - k_{0}\right)^{2}k_{0}^{2} + {1 \over 4}
\left(\Gamma_{\Delta} \left(\bar{k}\right)\right)^{2}} {
\left(p-k\right)^{4} \ln\left[-\left(p-k\right)^{2}\right] \over
\left(k_{0}^{2}-\bar{k}^{2} + i\epsilon\right)^{2}} \left[\left(2 p\cdot
k - k^{2}\right)\slash{k} - k^{2}\slash{p}\right]
\end{eqnarray*}
where $\zeta = {8 \pi^{3 \over 2} \left( A-1 \right) \rho_{0} 
f_{\pi}^{2} \omega_{\Delta} \over m_{\pi}^{2} c}$. The Borel transform 
with respect to $-p^2$, $\hat{B}$, eliminates terms proportional 
to $p \cdot k$ in the expansion of $\left(p-k\right)^{2n} 
\ln[-\left(p-k\right)^{2}]$, simplifying the integrand so that the 
term proportional to $\slash{k}$ vanishes from symmetry in the 
angle-averaging. We are left with integrals over the pion energy 
$k_{0}$ and the magnitude of the pion three-momentum $\bar{k}$. The 
integrand also picks up a Borel dependence of the form $M_{B}^{2n} 
\exp(-k^{2}/M_{B}^{2})$. The energy integration can be completed 
analytically, where we find that the factor $\exp(-k^{2}/M_{B}^{2}) 
\rightarrow 1$ and the dependence on the Borel mass can be taken out 
of the integral. This first diagram is found to be 
\begin{equation}
\hat{B} \Pi_{(4a,\rho)}(p)={\zeta i \over \lambda_{\pi}^{4}}I\left(
\alpha, c \right){1 \over \left(2 \pi \right)^{7}} \slash{p} M_{B}^{6}
\end{equation}
where $I\left( \alpha, c \right)$ is the remaining one-dimensional
integral over the pion three-momentum. Defining $\tilde{\Gamma}( \bar{k}
)= k_{0}\Gamma(\bar{k},k_{0})$, the integral takes the form
\begin{equation} \label{eq:uglyint}
I\left( \alpha, c \right)={- i \over 2 \pi^{2} c^{2}} \int_{0}^{\infty}
\!d\bar{k} F_{1}\left(\bar{k}, \alpha, c \right) \left( g \left( \bar{k}
\right) \right)^{2} {\omega_{\Delta} \bar{k} - \bar{k}^{2} + {i \over 2}
\tilde{\Gamma} \left( \bar{k} \right) \over \left(\omega_{\Delta}\bar{k}
- \bar{k}^{2}\right)^{2} + {1 \over 4}\left(\tilde{\Gamma}\left( \bar{k}
\right)\right)^{2}}. 
\end{equation} We calculate this integral numerically.

The other diagrams shown in Figure \ref{fig:medpi2dia} yield the
following contributions to the nucleon correlator:
\begin{eqnarray}
\hat{B} \Pi_{(4b,\rho)}(p)&=&-{\zeta i \over \lambda_{\pi}^{4}}I\left(
\alpha, c \right){ 4 \pi \over \left(2 \pi \right)^{6}} \left< \bar{q} q
\right> M_{B}^{4} \nonumber \\
\hat{B} \Pi_{(4c,\rho)}(p)&=&{\zeta i \over \lambda_{\pi}^{4}}I\left(
\alpha, c \right){ 2^{3} \pi \over \left(2 \pi \right)^{4}3} \slash{p}
\left< \bar{q} q \right>^{2} \nonumber \\
\hat{B} \Pi_{(4e,\rho)}(p)&=&-{\zeta i \over \lambda_{\pi}^{4}}I\left(
\alpha, c \right){ 1 \over 2^{9} \pi^{7}} \slash{p} \left< g_{s}^{2} G
\cdot G \right> M_{B}^{2} \nonumber \\
\hat{B} \Pi_{(4f,\rho)}(p)&=&-{\zeta i \over \lambda_{\pi}^{4}}I\left(
\alpha, c \right){ 1 \over 2^{4}3\pi^{7}} \slash{p} \left< \bar{q} q
\right>\left< \bar{q} g_{s} \sigma \cdot G q \right> {1 \over M_{B}^{2}}
\nonumber \\
\hat{B} \Pi_{(4g,\rho)}(p)&=&-{\zeta i \over \lambda_{\pi}^{4}}I\left(
\alpha, c \right){ 1 \over 2^{3}3^{2}\pi^{5}} \left< \bar{q} q
\right>\left< g_{s}^{2} G \cdot G \right> \nonumber \\
\hat{B} \Pi_{(4h,\rho)}(p)&=&{\zeta i \over \lambda_{\pi}^{4}}I\left(
\alpha, c \right){ 17 \over 3^{4}\pi^{3}} g_{s}^{2} \left< \bar{q} q
\right>^{3}{1 \over M_{B}^{2}},
\end{eqnarray} where $I\left( \alpha, c \right)$ in each of these
expressions is the same as Equation (\ref{eq:uglyint}).

\section{Construction of the In-Medium Sum Rules}

\subsection{In the Chiral Limit}

The phenomenological (``right-hand'') side of the sum rules, is
expressed as a dispersion relation with a lowest-lying hadronic pole of
in-medium mass $m^{*}$ and residue $\lambda_{p}^{*2}$ clearly separated
from a continuum of higher-energy states with threshold
$s_{0}^{*}$\cite{druklev}:
\begin{equation}
\Pi^{p}_{A,{\text {RHS}}}(p)=\lambda_{p}^{*2} c_{1}^{2} { \slash{p} +
m^{*} \over p^{2} - m^{*2}} + {1 \over \pi}\int_{s_{0}^{*}}^{\infty}\!ds
{{\text {Im}}\Pi_{A}^{{\text {cont}}}\left( s, u \cdot q \right) \over s
- p^{2}}
\end{equation} 
Guided by the vacuum calculations, we assume that the structure
parameter associated with the current including the pion cloud
$\lambda_{p}^{'*2} \ll \lambda_{p}^{*2}$, so there is no additional pole
considered in our model of the phenomenological correlator. Rather than
find the values of the parameters in medium, we follow the appoach of
Drukarev and Levin and determine the shifts of the in-medium parameters
from their vacuum values\cite{druklev}. We remove the contributions to
the nucleon correlator due to the pure in-vacua diagrams by defining the
quantity
\begin{equation}
\phi^{(i)} \equiv -2 i \left( 2 \pi \right)^{4} \hat{B} \left[
\Pi_{A}^{(i)}\left(p \right) - \Pi^{(i)}\left(p \right)\right],
\end{equation} 
where $i$ labels the three structures in the sum rule (${\mathbf 1}$,
$\slash{p}$, and $\slash{u}$, respectively). On the phenomenological
side of the sum rule we can express $\phi^{(i)}$ in terms of the
parameter shifts and the Borel transform of the RHS in vacuum as
\begin{equation}
\phi^{(i)}\left(M_{B}^{2},s\right)= \left[\Delta m{\partial \ \over
\partial m} +  \Delta \tilde{\lambda}^{2}{\partial \ \  \over \partial
\tilde{\lambda}^{2} } + \Delta s_{0} {\partial \ \  \over \partial
s_{0}}\right] \hat{B} \Pi_{0,{\text {RHS}}}^{(i)}\left( p \right)
\end{equation} 
for $i=s,p$ (since there is no $u$ structure in vacuum). The
contributions to the microscopic side of the sum rule are given in the
previous section. With the subscript labelling the contributions due to
each figure, we find in the nuclear rest frame using the constraint in
Equation (\ref{eq:constraint}):
\begin{mathletters}
\label{eq:phimicro}
\begin{eqnarray} 
\phi_{2}^{(s)}\left(M_{B}^{2}\right)&=&2 \rho a_{N} M_{B}^{4} E_{1} +
\left[ - 3 \cdot 2^{3} \pi^{2}  \rho m^{2} a_{0} - {2^{2} \over 3^{2}}
\rho \left( b_{N} a_{0} + b_{0}a_{N}\right)\right] + {2 \cdot 17 \over
3^{2} \pi^{2}}g_{s}^{2}\rho a_{0}^{3} {1 \over M_{B}^{2}} \nonumber \\
\phi_{2}^{(p)}\left(M_{B}^{2}\right)&=&2^{4}\pi^{2} \rho M_{B}^{4} E_{1}
L^{-{4 \over 9}} + \left[ {1 \over 2^{2}}\rho b_{N} - {2^{4} \pi^{2}
\over 3}m^{2} \rho\right] M_{B}^{2}E_{0}L^{- {4 \over 9}} + \left[ 
{2^{3} \over 3}\rho a_{0}a_{N} L^{{4 \over 9}} - {\pi^{2} \over 3 \cdot
2} \rho b_{0} L^{-{4 \over 9}}\right] \nonumber \\ 
& & - \pi^{2} \rho b_{0}m^{2} 
      L^{-{4 \over 9}}{1 \over M_{B}^{2}} \nonumber \\
\phi_{2}^{(u)}\left(M_{B}^{2}\right)&=&-2^{6}\pi^{2}\rho M_{B}^{4} E_{1}
L^{-{4 \over 9}} - {5 \pi^{2} \over 3 \cdot 2} \rho b_{0} 
L^{-{4 \over 9}} \\
\phi_{3}^{(s)}\left(M_{B}^{2}\right)&=& 0\nonumber \\
\phi_{3}^{(p)}\left(M_{B}^{2}\right)&=& {1 \over \lambda_{\pi}^{4}}
\left( -{3 \over 2^{2}}\rho M_{B}^{8} E_{3} L^{-{4 \over 9}} + \left[ {3
\over 2^{2}} m^{2}\rho - {1 \over 2^{7} \pi^{2}} \rho b_{N} \right]
M_{B}^{6}E_{2} L^{-{4 \over 9}} \right. \nonumber \\
& &\left. +\left[-{1 \over 2 \cdot 3 \pi^{2}}\rho a_{0}a_{N} L^{{4 \over
9}} - {11 \pi^{4} \over 2^{5} 3^{2}}\rho b_{0}L^{-{4 \over
9}}\right]M_{B}^{4}E_{1} + {11 \pi^{4} \over 2^{5} 3^{2}} \rho b_{0}
M_{B}^{2} E_{0} L^{-{4 \over 9}} \right) \nonumber \\
\phi_{3}^{(u)}\left(M_{B}^{2}\right)&=& {1 \over \lambda_{\pi}^{4}}
\left({3 \over 2} \rho M_{B}^{8} E_{3} L^{-{4 \over 9}} + {5 \over 2^{3}
3^{2}} \rho b_{0} M_{B}^{4} E_{1} L^{-{4 \over 9}} \right) \\
\phi_{4}^{(s)}\left(M_{B}^{2}\right)&=&{\zeta \over \lambda_{\pi}^{4}}
{\text {Re}}\left[ I\left(\alpha, c\right) \right] \left[{a_{0} \over
\pi^{3}} M_{B}^{4} E_{1} + {2 a_{0} b_{0} \over 3^{2} \pi^{3}}  - {17
\over 3^{4} \pi^{5}}g_{s}^{2} a_{0}^{3} {1 \over M_{B}^{2}}\right]
\nonumber \\
\phi_{4}^{(p)}\left(M_{B}^{2}\right)&=& {\zeta \over \lambda_{\pi}^{4}}
{\text {Re}}\left[ I\left(\alpha, c\right) \right] \left[ {1 \over 2
\pi^{3}} M_{B}^{6} E_{2} L^{-{4 \over 9}} - {b_{0} \over 2^{3} \pi^{3}}
M_{B}^{2} E_{0} L^{-{4 \over 9}} + {2 a_{0}^{2} \over 3 \pi^{3}}L^{{4
\over 9}} - {a_{0}^{2} m_{0}^{2} \over 3 \cdot 2^{2} \pi^{7}}  {1 \over
M_{B}^{2}} \right] \nonumber \\
\phi_{4}^{(u)}\left(M_{B}^{2}\right)&=& 0.
\end{eqnarray}
\end{mathletters} Here we have used for the condensates
\begin{eqnarray*}
a_{0}&=&-\left(2 \pi\right)^{2}\left<\bar{q}q\right>
 =0.55 \ \text{GeV}^{3},\\ 
a_{N}&=&-\left(2 \pi\right)^{2}
\left<N\right|\bar{q}q\left|N\right>, \\
b_{0}&=&\left<g_{s}^{2}G \cdot G\right>
 =0.474 \ \text{GeV}^{4}, \\
b_{N}&=&\left<N\right|g_{s}^{2}G \cdot G\left|N\right>, 
\text{ and} \\
m_{0}^{2}&=&{\left<\bar{q} g_{s} \sigma \cdot G q\right> 
\over a_{0}}= 0.8 \ \text{GeV}^{2}.
\end{eqnarray*}
The left-hand side of the sum rule is then  $\phi^{(i)} = c_{1}^{2}
\phi^{(i)}_{2} - c_{2}^{2} \phi^{(i)}_{3} - c_{2}^{2} \phi^{(i)}_{4}$.
We have taken into account perturbative corrections using the
renormalization group anomalous dimensions of the interpolating field
and the local operators used in constructing the LHS. Each term is
multiplied by the appropriate power of the anomalous dimension
factor\cite{ioffe,is} \begin{equation}
L\equiv {\ln\left( M_{B} / \Lambda_{{\text {QCD}}}\right) \over
\ln\left(\mu / \Lambda_{{\text {QCD}}}\right)}.
\end{equation} 
Since it arises from a conserved current, the anomalous dimension of the
pseudo-Goldstone boson is zero. We set the renormalization point of the
OPE $\mu = 500 \ {\text {MeV}}$ and the QCD scale parameter
$\Lambda_{{\text {QCD}}} = 100 \ {\text {MeV}}$. In the continuum model,
the contributions from higher-energy states are included by using the
prescription whereby terms proportional to $M_{B}^{2n}$ are multiplied
by factors of $E_{n-1}$ defined as
\begin{equation}
E_{n} \equiv 1 - \exp\left(-s_{0}^{*} / M_{B}^{2}\right)\sum_{k=0}^{n}
{1 \over k!} \left({s_{0}^{*} \over M_{B}^{2}}\right)^{k}.
\end{equation} 

We introduce a pair of functions $\Phi^{(i)}$ as
\begin{eqnarray}
\Phi^{(s)} \left(M_{B}^{2},s\right) & = &
\phi^{(s)}\left(M_{B}^{2},s\right) \nonumber \\ 
& = & c_{1}^{2} \left\{ \tilde{\lambda}^{2} \left( 1 - {2 m^{2} \over
M_{B}^{2}} \right) e^{-{m^{2} \over M_{B}^{2}}} \Delta m + m e^{-{m^{2}
\over M_{B}^{2}}} \Delta \tilde{\lambda}^{2} - 2 a s_{0}e^{-{s_{0} \over
M_{B}^{2}}} \Delta s_{0} \right\}\\
\Phi^{(p)} \left(M_{B}^{2},s\right) & = & -m
\phi^{(p)}\left(M_{B}^{2},s\right), \nonumber \\
& = & c_{1}^{2} \left\{{2 m^{2} \over M_{B}^{2}} \tilde{\lambda}^{2}
e^{-{m^{2} \over M_{B}^{2}}} \Delta m  - m e^{-{m^{2} \over M_{B}^{2}}}
\Delta \tilde{\lambda}^{2} \right. \nonumber \\
& & \left. + {m \over 2}  L^{-{4 \over 9}} \left[ {s_{0}^{4} \over
\lambda_{\pi}^{4}  3 \cdot 2^{5}5} + s_{0}^{2}\left( 1 - {b \over
\lambda_{\pi}^{4} 3 \cdot 2^{5}} \right) + {a^{2} L^{{8 \over 9}} s_{0}
\over \lambda_{\pi}^{4} 3 \cdot 2^{2}} + \left({b \over 2} + {m_{0}^{2}
a^{2} \over \lambda_{\pi}^{4} 3 \cdot 2^{2}}\right) \right] e^{{-s_{0}
\over M_{B}^{2}}} \Delta s_{0} \right\} .
\end{eqnarray}
The sum of these functions has no dependence on the residue shift
$\Delta\tilde{\lambda}^{2}$, and sufficient cancellation occurs between
the terms proportional to the threshold shift $\Delta s_{0}$ that we can
ignore these terms. The in-nucleus mass shift of the nucleon can then be
determined using
\begin{equation}
\Delta m = {1 \over c_{1}^{2} \tilde{\lambda}^{2}} \exp\left({m^{2}
\over M_{B}^{2}}\right)\left[\Phi^{(s)} \left(M_{B}^{2},s\right) +
\Phi^{(p)} \left(M_{B}^{2},s\right) \right].
\end{equation}

We adopt a derivative improvement to the sum rules suggested by Drukarev
and Ryskin \cite{drukrysk} where
\begin{equation}
\psi_{j}^{(i)} \equiv {M_{B}^{4} \over m^{2}} {d \phi_{j}^{(i)} 
\over dM_{B}^{2}}.
\end{equation} 
This improvement minimizes the effect of the factorization approximation
of the four quark condensates by eliminating the terms proportional to
$p^{-2}$ from the dominant contributing diagrams, therefore aiding the
convergence of the sum rules. Applying the derivative operator to
Equations (\ref{eq:phimicro}) we find
\begin{mathletters}
\begin{eqnarray} \label{eq:psimicro}
\psi_{2}^{(s)}\left(M_{B}^{2}\right)&=&2^{2} \rho a_{N} {1 \over m^{2}}
M_{B}^{6} E_{2} - {1 \over m^{2}} {2 \cdot 17 \over 3^{3}
\pi^{2}}g_{s}^{2}\rho a_{0}^{3} \nonumber \\
\psi_{2}^{(p)}\left(M_{B}^{2}\right)&=&2^{5}\pi^{2} \rho {1 \over m^{2}}
M_{B}^{6} E_{2} L^{-{4 \over 9}} + \left[ {1 \over 2^{2}}\rho b_{N}{1
\over m^{2}} - {2^{4} \pi^{2} \over 3} \rho\right] M_{B}^{4}E_{1}L^{- {4
\over 9}} + \pi^{2} \rho b_{0}L^{-{4 \over 9}} \nonumber \\
\psi_{2}^{(u)}\left(M_{B}^{2}\right)&=&-2^{7}\pi^{2}\rho {1 \over m^{2}}
M_{B}^{6} E_{2} L^{-{4 \over 9}} \\
\psi_{3}^{(s)}\left(M_{B}^{2}\right)&=& 0\nonumber \\
\psi_{3}^{(p)}\left(M_{B}^{2}\right)&=& {1 \over \lambda_{\pi}^{4}}
\left( -{3 \over m^{2}}\rho M_{B}^{10} E_{4} L^{-{4 \over 9}} + \left[
{3^{2} \over 2^{2}} \rho - {3 \over 2^{7} \pi^{2}} {\rho b_{N} \over
m^{2}} \right] M_{B}^{8}E_{3} L^{-{4 \over 9}} \right. \nonumber \\
& &\left. +\left[-{1 \over 3 \pi^{2}}\rho {a_{0}a_{N} \over m^{2}} L^{{4
\over 9}} - {11 \pi^{4} \over 2^{4} 3^{2}}{\rho b_{0}\over m^{2}}L^{-{4
\over 9}}\right]M_{B}^{6}E_{2} + {11 \pi^{4} \over 2^{5} 3^{2}} {\rho b_{0}
\over m^{2}} M_{B}^{4} E_{1} L^{-{4 \over 9}} \right)\nonumber \\
\psi_{3}^{(u)}\left(M_{B}^{2}\right)&=& {1 \over \lambda_{\pi}^{4}}
\left({3 \cdot 2 \over m^{2}} \rho M_{B}^{10} E_{4} L^{-{4 \over 9}} +
{5 \over 2^{3} 3^{2}} \rho {b_{0} \over m^{2}} M_{B}^{6} E_{2} L^{-{4
\over 9}} \right) \\
\psi_{4}^{(s)}\left(M_{B}^{2}\right)&=&{\zeta \over \lambda_{\pi}^{4}}
{\text {Re}}\left[ I\left(\alpha, c\right) \right]  \left[{2 a_{0} \over
m^{2} \pi^{3}} M_{B}^{6} E_{2} +  {17 \over 3^{4} \pi^{5}}{g_{s}^{2}
\over m^{2}} a_{0}^{3}\right] \nonumber \\
\psi_{4}^{(p)}\left(M_{B}^{2}\right)&=& {\zeta \over \lambda_{\pi}^{4}}
{\text {Re}}\left[ I\left(\alpha, c\right) \right] \left[ {3 \over 2
\pi^{3}} {1 \over m^{2}} M_{B}^{8} E_{3} L^{-{4 \over 9}} - {b_{0} \over
2^{3} \pi^{3}}{1 \over m^{2}} M_{B}^{4} E_{1} L^{-{4 \over 9}} +
{a_{0}^{2} m_{0}^{2} \over 3 \cdot 2^{2} \pi^{7}}  {1 \over m^{2}}
\right] \nonumber \\
\psi_{4}^{(u)}\left(M_{B}^{2}\right)&=& 0
\end{eqnarray}
\end{mathletters}
The improvement does not remove all dependence of the LHS on the
four-quark condensate, but these remaining four-quark terms are largely
suppressed in comparison to terms with similar Borel weighings. As with
the unimproved sum rules, a linear combination of these expressions can
be built that has no dependence on the residue shift and negligible
dependence on the threshold shift, giving the mass shift of the nucleon
as
\begin{equation} 
\label{eq:psimassshift}
\Delta m = {1 \over c_{1}^{2}\tilde{\lambda}^{2}} \exp\left({m^{2} \over
M_{B}^{2}}\right)\left[\psi^{(s)} \left(M_{B}^{2},s\right) - m
\psi^{(p)} \left(M_{B}^{2},s\right) \right].
\end{equation} 
The improved sum rule is used in the analysis of the nucleon mass shifts
in this work.

\subsection{Beyond the Chiral Limit}

As discussed above, in Ref. \cite{kissl1} it was shown that processes
illustrated in Figure \ref{fig:chivac} produce terms proportional to
$m_{\pi}^{3} / f_{\pi}^{2}$. Also the $m_\pi^2 ln(m_\pi^2)$ terms were
shown to cancel using the analysis of Ref. \cite{lccg}. It was thereby
shown that our pion cloud model is consistent with chiral perturbation
theory to leading order in the nonanalytic contributions to the nucleon
mass in vacuum, and that we could use the chiral perturbation theory
result to constrain the parameters of the model

\begin{figure} 
\begin{center}
\epsfxsize=125pt \epsfbox{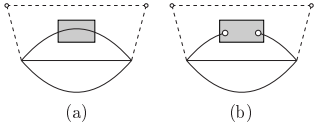}
\end{center}
\caption{The diagrams corresponding to the lowest-dimension chiral
corrections to the nucleon correlator in medium.} 
\label{fig:chidi}
\end{figure}

In dense nuclear matter (and at finite temperature, Refs.
\cite{eletsky,koike}) another problem with chiral symmetry breaking
seems to occur in the QCD sum rule method \cite{birse}. Equation
(\ref{eq:qcondm}) can be rewritten as
\begin{equation}
{\left<\bar{q}q\right>_{\rho} \over
\left<\bar{q}q\right>_{0}}=-\rho{\sigma_{\pi N} \over f_{\pi}^{2} 
m_{\pi}^{2}}.
\end{equation}
However, the LNAC to the $\pi N$ sigma commutator is proportional to
$m_{\pi}^{3}$ \cite{gasser2}). The leading nonanalytic contribution to
the in-medium quark condensate is thereby seen to be of order $\rho
m_{\pi}$, which is forbidden by chiral counting rules \cite{weinberg}.
Birse and Krippa accounted for this by adding contributions to the RHS
where the nucleon current interacts with soft pions from the pion cloud
of another nucleon in the nucleus \cite{birse}. Using the soft pion
theorem of PCAC \cite{pcac} they find a term proportional to $\rho
m_{\pi}$ on the phenomenological side which cancels the offending term
on the microscopic side arising from the in-medium quark condensate. As
we go away from the chiral limit in our pion cloud model, we adopt the
same modification to the phenomenological nucleon correlator, and in
doing so we need to make sure that there are no contributions to the OPE
from the pion cloud that would violate the chiral counting rules to
leading nonanalytic order. This is indeed the case, since the chiral
expansion of the pion propagator contributes lowest order terms
proportional to $m_{\pi}^{2}$, maintaining the leading order behavior.
Likewise, similar to the vacuum contribution illustrated in Figure
\ref{fig:chivac}a lowest dimension in-medium contributions shown in
Figures \ref{fig:chidi}a and \ref{fig:chidi}b are possible. However,
their effects are on higher order nonanalytic terms as they provide
corrections proportional to $\rho m_{\pi}^{3}$ and $\rho
m_{\pi}^{4}\ln(-m_{\pi}^{2})$. Thus, to first order in baryon density
the pion cloud model for the nucleon correlator in a nucleus is
consistent with chiral perturbation theory in the leading nonanalytic
contributions to the nucleon mass.

\section{Results for the Nucleon Mass Shift in Spin-Isospin Symmetric Nuclei}

This model is best applied to light nuclei where there are large surface
volume regions and therefore the density gradients extend over a large
fraction of the nucleus. In infinite nuclear matter these denisty
gradients are zero, so that the s-wave part of the $\pi A$ optical
potential, which we neglect dominates the in-medium pion cloud
contribution to the nucleon correlator. In matter with $Z=N$, the
strength of this potential is very small and the contributions of the
diagrams in Figure \ref{fig:medpi2dia} to the nucleon mass are
negligible. As a result, the pion cloud has almost no effect on the
nucleon mass in infinite nuclear matter, similar to the pion cloud
effect on the nucleon mass in vacuum.  It is evident that in large
nuclei the pion cloud contribution to the nucleon mass becomes
non-negligible only near the nuclear surface, a well-known property of
the Kisslinger potential. So long as the nucleon remains  a distance
away from the surface comparable to about twice the $\pi N$ interaction
range (with $r_{\pi N} \sim 0.7 \ {\text {fm}}$) our mass shift result
for this nucleon reproduce the work of previous authors
\cite{druklev,drukrysk}. Also  application to large nuclei would require
an extension of the method to account for $N \neq Z$.

In this paper, we work with models of two small symmetric nuclei that
are familiar from $\pi A$ scattering experiments, $\,^{16}{\text {O}}$
and $\,^{40}{\text {Ca}}$. The nucleon density distribution of the
ground states of both nuclei are given by Eq (\ref{eq:density})). For
$\,^{16}{\text {O}}$ we choose the parameters $c=1.75 \ {\text {fm}}$
and $\alpha=2.0$ \cite{dist}. This produces a distribtuion of nucleons
with a local minimum at $r=0$ and an RMS radius of $\sim 2.6 \ {\text
{fm}}$. The parameters for the $\,^{40}{\text {Ca}}$ density
distribution are taken as $c=2.80 \ {\text {fm}}$ and $\alpha=1.1$.

\begin{figure} 
\begin{center} 
\epsfbox{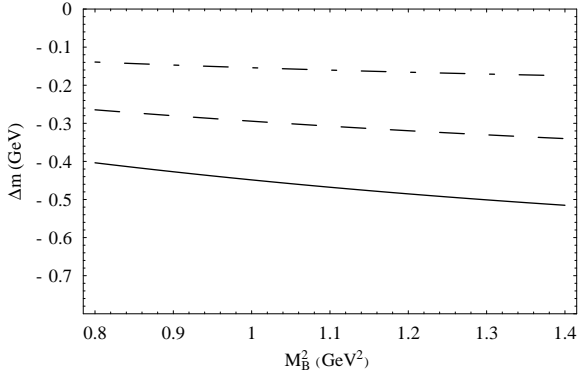} 
\end{center}
\caption{The mass shifts in the $\,^{16}{\text {0}}$ nucleus at
$0.5\rho_{0}$ with constant pion cloud probabilities. The solid line is
the total mass shift, the dashed line is the mass shift due to the
correlator without the pion cloud, and the dot-dashed line is the shift
due to the pion cloud contribution.} 
\label{fig:Osurfcon} 
\end{figure}

\begin{figure} 
\begin{center} 
\epsfbox{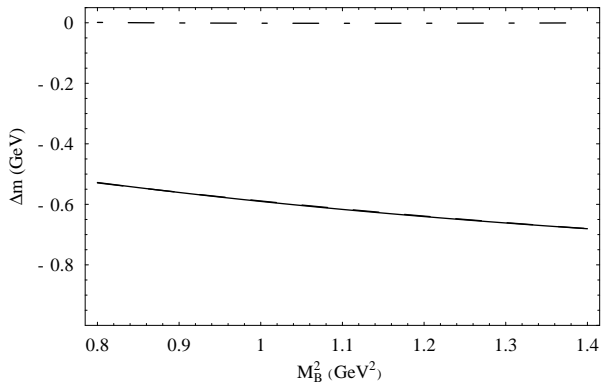} 
\end{center}
\caption{The mass shifts in the $\,^{16}{\text {0}}$ nucleus at
$\rho_{0}$ with constant pion cloud probabilities. The solid line is the
total mass shift, the dashed line is the mass shift due to the
correlator without the pion cloud, and the dot-dashed line is the shift
due to the pion cloud contribution.} 
\label{fig:Ocentcon} 
\end{figure}

\begin{figure}
\begin{center}
\epsfbox{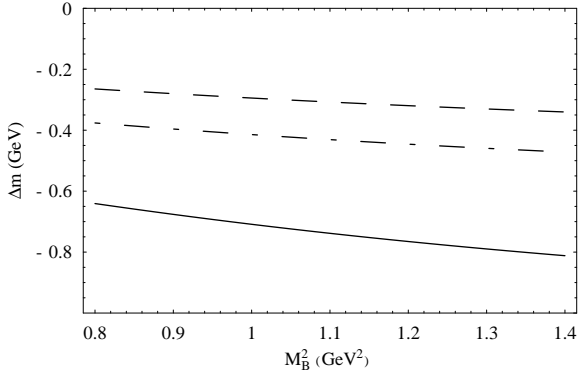}
\end{center}
\caption{The mass shifts in the $\,^{40}{\text {Ca}}$ nucleus at
$0.5\rho_{0}$ with constant pion cloud probabilities. The solid line is
the total mass shift, the dashed line is the mass shift due to the
correlator without the pion cloud, and the dot-dashed line is the shift
due to the pion cloud contribution.}
\label{fig:Casurfcon}
\end{figure}

\begin{figure}
\begin{center}
\epsfbox{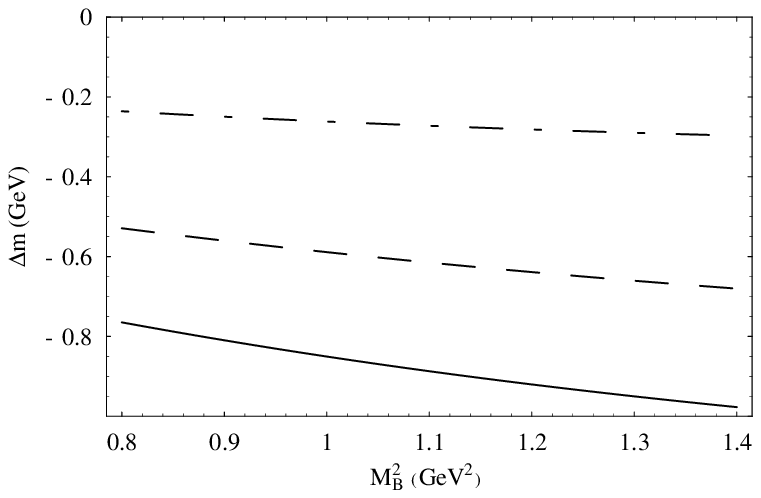}
\end{center}
\caption{The mass shifts in the $\,^{40}{\text {Ca}}$ nucleus at
$\rho_{0}$ with constant pion cloud probabilities. The solid line is the
total mass shift, the dashed line is the mass shift due to the
correlator without the pion cloud, and the dot-dashed line is the shift
due to the pion cloud contribution.}
\label{fig:Cacentcon}
\end{figure}

We find the mass shifts in the chiral limit, ignoring in our analysis
the effects of the small leading nonanalytic terms discussed above. The
mass shift of a nucleon in each nucleus is found by evaluating Eq.
(\ref{eq:psimassshift}) in a Borel window $0.8 \ {\text {GeV}} \leq
M_{B}^{2} \leq 1.4 \ {\text {GeV}}$. If the plot of the mass shift as a
function of the Borel mass is relatively flat in this window, then the
sum rules are sufficiently convergent and the mass shift is taken at
this flat value. To test the stability of the sum rules we look at the
mass shift at two points in the nucleus: first in the middle of the
surface where the local density is half the nuclear matter density
$\rho_{0}$, then at the center of the nucleus, where the central density
is equivalent to $\rho_{0}$. As can be seen in Figures
\ref{fig:Osurfcon}-\ref{fig:Cacentcon}, the sum rules in both nuclei
provide sufficiently stable solutions in the Borel window.

\newpage

\begin{figure}
\begin{center}
\epsfbox{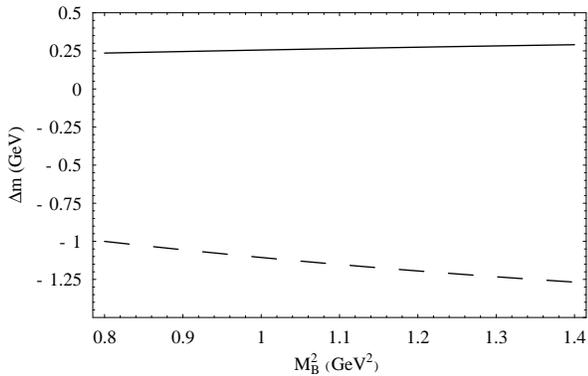}
\end{center}
\caption{The mass shift from the correlator structures in the
$\,^{40}{\text {Ca}}$ nucleus at $\rho_{0}$ with constant pion cloud
probabilities. The solid line is the mass shift from vector
contributions to the correlator, and the dashed line is the mass shift
due to scalar contributions to the correlator.}
\label{fig:Cavecsca}
\end{figure}
The convergence is amplified in the plots of the individual
contributions of the vector and scalar correlators (Fig.
\ref{fig:Cavecsca}). We find that the nucleon mass shift in an
$\,^{16}{\text {O}}$ nucleus is negligible at the central baryon
density, and $-150 \ {\text {MeV}}$ at one-half central density. The
mass shifts in a $\,^{40}{\text {Ca}}$ nucleus are $-260 \ {\text
{MeV}}$ at $\rho = \rho_{0}$ and $-400 \ {\text {MeV}}$ at $\rho =
\rho_{0}/2$.

\begin{figure}
\begin{center}
\epsfbox{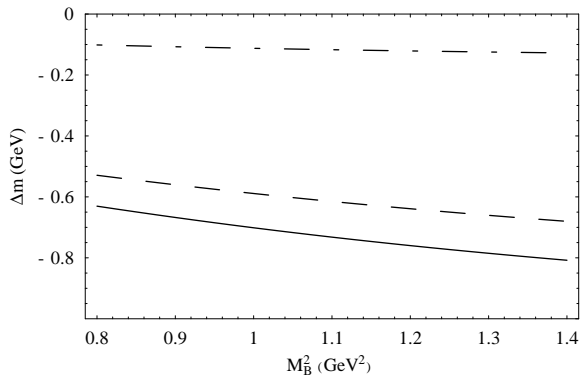}
\end{center}
\caption{The mass shifts in the $\,^{40}{\text {Ca}}$ nucleus at
$\rho_{0}$ with pion cloud probabilities linearly scaling with density.
The solid line is the total mass shift, the dashed line is the mass
shift due to the correlator without the pion cloud, and the dot-dashed
line is the shift due to the pion cloud contribution.}
\label{fig:Cacentsca}
\end{figure}

Some pion cloud effects might be related to the spontaneous breaking of
chiral symmetry. Since  the reduction in the value the quark condensate
is a signal of partial restoration of chiral symmtery in the nucleus,
the pion cloud amplitude $c_{2}$ might decrease accordingly. We explore
this idea by  modifying our model, giving the pion cloud probability
$c_{2}^{2}$ a linear dependence on density such that it has the vacuum
value at zero density, and it vanishes at the density where the quark
condensate $\left< \bar{q}q \right>_{\rho}$ goes to zero. Similarly, the
probability of no pion cloud $c_{1}^{2}$ increases from the vacuum value
of $0.5$ to the phase transition point value of $1.0$. Figure
\ref{fig:Cacentsca} shows the mass shift of a nucleon in a
$\,^{40}{\text {Ca}}$ nucleus at central density utilizing this
modification to the model. The shift in $\,^{16}{\text {O}}$ shows a
similar, if less dramatic, reduction. As in the constant pion cloud
probability calculation, the sum rule calculation is sufficiently flat
in the Borel window to provide a value for the mass shift.  The mass
shifts in $\,^{16}{\text {O}}$ are $\sim 0 \ {\text {MeV}}$ at central
and $-100 \ {\text {MeV}}$ at surface density. In $\,^{40}{\text {Ca}}$
we find $-100 \ {\text {MeV}}$ and $-275 \ {\text {MeV}}$, respectively.

\begin{figure}
\begin{center}
\epsfbox{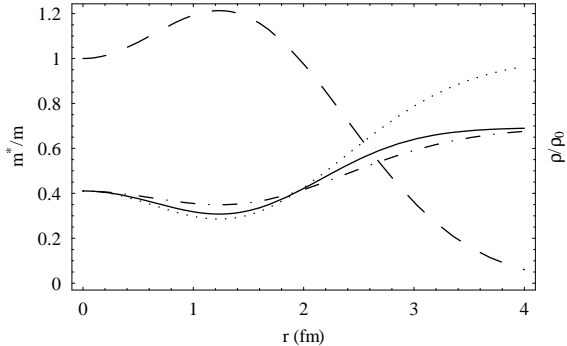}
\end{center}
\caption{The mass shifts in the $\,^{16}{\text {O}}$ nucleus as
functions of distance from the nuclear center, calculated at a Borel
mass of 1 GeV. The dashed line is the density distribution. The solid
line is the nucleon mass with linearly scaling pion cloud probabilities,
the dot-dashed line is the mass with constant probabilities, and the
dotted line is the nucleon mass without the pion cloud contributions.}
\label{fig:Ofunrad}
\end{figure}

\begin{figure}
\begin{center}
\epsfbox{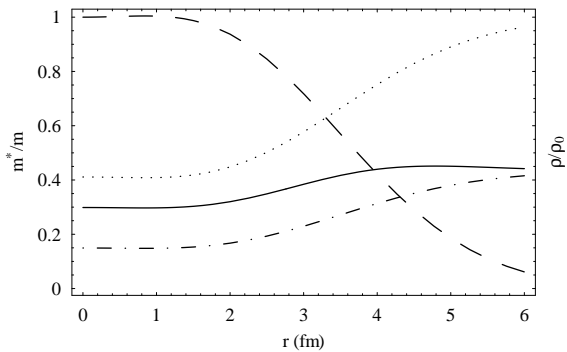}
\end{center}
\caption{The mass shifts in the $\,^{40}{\text {Ca}}$ nucleus as
functions of distance from the nuclear center, calculated at a Borel
mass of 1 GeV. The dashed line is the density distribution. The solid
line is the nucleon mass with linearly scaling pion cloud probabilities,
the dot-dashed line is the mass with constant probabilities, and the
dotted line is the nucleon mass without the pion cloud contributions.}
\label{fig:Cafunrad}
\end{figure}

We compare these two methods to the sum rules without an explicit pion
cloud throughout the nucleus in Figures \ref{fig:Ofunrad} and
\ref{fig:Cafunrad}. In these calculations, the density used at each
point in the sum rules is the local density obtained from the density
distribution (Eq. (\ref{eq:density})) and the nucleon mass is plotted as
a function of the distance from the center of the nucleus. Each mass
shift is evaluated at the Borel mass $M_{B}=1.0 \ {\text {GeV}}$.

In an $\,^{16}{\text {O}}$ nucleus, the mass shift due to the pion cloud
is small compared to the mass shift from the non-pionic terms but
positive out to a distance of about $4.5 \ {\text {fm}}$ from the
nuclear center. As the nucleon appraoches the nuclear surface the pion
cloud contribution to the mass shift becomes positive and larger than
the mass shift without pion cloud. Even though the local density in the
interval $3.0 \ {\text {fm}} \leq r \leq 4.0 \ {\text {fm}}$ is small
(hence the small mass shift due to the correlator contributions without
the pion cloud), the pion cloud is still sampling a significant range of
density gradients leading to a sizable optical potential strength, and
therefore a large mass shift. The two different approaches to the pion
cloud amplitudes in $\,^{16}{\text {O}}$ do not yield significantly
different results throughout the nucleus. In the $\,^{40}{\text {Ca}}$
nucleus, the mass shift due to the pion cloud is always negative and
comparable to the non-pionic mass shift. As with the oxygen nucleus, the
largest mass shift from the pionic terms occurs near the surface,
approaching $-600 \ {\text {MeV}}$. The constant and linearly decreasing
pion cloud probability calculations converge to the same value in this
region. The nuclear interior shows a significant difference between the
three methods. If $c_{1}$ and $c_{2}$ remain constant over the range of
densities, then the pion cloud provides a rather large mass shift ($\sim
-275 \ {\text {MeV}}$ at central density). Once $c_{1}$ and $c_{2}$ are
dependent on density, the pion cloud contribution is more than halved at
the center of the nucleus, with a mass shift closer to $-100 \ {\text
{MeV}}$. Clearly the choice of pion cloud method has a significant
effect in larger nuclei.

\section{Conclusions}

We have described a model based on QCD for the nucleon in a finite spin
and isospin symmetric nucleus including a pseudo-Goldstone boson
component of the correlator This approach is based on the model proposed
by one of us in Ref.\cite{kissl1}, with the extension of the pion free
propagator to the in-medium propagator using an optical potential which
is first order in the nuclear density. A fit to chiral perturbation
theory restricts the parameters of the pion cloud component. Our main
conclusion is that there are significant contributions to the mass of
the nucleon in the nucleus arising from the pion cloud component of the
nucleon.

In the deep interior of the oxygen nucleus we find a small positive mass
shift which becomes large and negative near the surface. In the calcium
nucleus, however, the mass shift is everywhere large and negative. We
introduce a parametrization of the pion cloud probability $c_{2}^{2}$
which decreases linearly from the vacuum value to zero at the chiral
symmetry restoration as a possible test of partial restoration of chiral
symmetry. At high densities, the effect of the pion cloud is then
reduced leading to smaller mass shifts.

One obvious extension to this model is the consideration of the strange
sector. By allowing for the inclusion of kaon cloud terms in the baryon
current, we can construct similar sum rules for the mass shift of
hyperons.  The primary obstacles are the uncertainties of several
in-medium matrix elements on the microscopic side, the complicated
structure of the dispersion relation on the phenomenological side
\cite{henpas}, and the difficulty in finding an appropriate form for the
$K$-$A$ optical potential to determine the kaon self-enegry. The model
can also be modified to handle N $\neq$ Z nuclei. The nucleon correlator
in such nuclei without the pion cloud has been discussed in
Ref.\cite{druklev}, and the optical potential is easily treated in
asymmetric nuclei by including the isospin-dependent and spin flip terms
in the $\pi N$ scattering amplitude (Eq. (\ref{eq:pinf})). A third venue
of further research concerns the applications of the model to sea quark
distributions in nuclei and the development of an improvement to QCD
calculations of nuclear deep inelastic scattering in the interval
$0.1\leq x \leq 0.3$.

\vspace{5 mm}

We would like to thank Mountaga Aw and Otto Linsuain for helpful 
discussions.


\begin{references}
\bibitem{druklev}E. G. Drukarev and E. M. Levin, JETP Lett. {\bf 48},
338 (1988); Sov. Phys. JETP {\bf 68}, 680 (1989); Nucl. Phys. {\bf
A511}, 679 (1990); Progr. Part. Nucl. Phys. {\bf 27}, 77 (1991).
\bibitem{bethe}H. A. Bethe, Ann. Rev. Nucl. Sci. {\bf 21}, 125 (1971).
\bibitem{qhd}J. D. Walecka, ``Theoretical Nuclear and Subnuclear
Physics'' (Oxford University Press, New York) (1995). 
\bibitem{lp}P. Langacker and H. Pagels, Phys. Rev. {\bf D8}, 
4595 (1973).
\bibitem{gl}J. Gasser and H. Leutwyler, Ann. Phys. {\bf 158}, 142
(1984). 
\bibitem{kissl1}L. S. Kisslinger, {\tt hep-th/9811497}.
\bibitem{nmc}P. Amadruz, et. al., Phys. Rev. Lett. {\bf 66}, 2717
(1991); \\ M. Arnedo, et.al., Phys. Rev {\bf D50}, R1 (1994).
\bibitem{e866}E. A. Hawker, et. al., Phys. Rev. Lett. {\bf 80}, 3715
(1998) ; \\ J. C. Peng, et. al., Phys. Rev. {\bf D56}, 92004 (1998).
\bibitem{pqcd}D. A. Ross and C. T. Sachrajda, Nucl. Phys. {\bf B149},
497 (1979). 
\bibitem{nqcd}H. Jung and L. S. Kisslinger, Nucl. Phys. {\bf
A586}, 682 (1995). \bibitem{sul}J. D. Sullivan, Phys. Rev. {\bf D5},
1732 (1972). 
\bibitem{st} J. Speth and A. W. Thomas, Adv. Nucl. Phys.
{\bf 24}, 83 (1998). \bibitem{mmp}P. L. McGaughey, J. M. Moss and J. C.
Peng, Ann. Rev. Nucl. Part. Sci. {\bf 49}, 217 (1999). 
\bibitem{gz} J. Gasser and A. Zepeda, Nucl. Phys. {\bf B174}, 445 (1980). 
\bibitem{lccg} S. H. Lee, S. Choe, T. D. Cohen, and D. K. Griegel, 
Phys. Lett. {\bf B348}, 263 (1995). 
\bibitem{birse}M. C. Birse, Phys. Rev. {\bf C 53}, R2048 (1996);\\  
M. C. Birse and B. Krippa, Phys. Lett. {\bf B 381}, 397 (1996). 
\bibitem{ioffe}B. L.Ioffe, Nucl. Phys. {\bf B188}, 317 (1981);
{\bf B191}, 591 (E) (1981);\\ V. M. Belayev and B. L. Ioffe, Sov. Phys.
JETP {\bf 56}, 493 (1982). 
\bibitem{kissl2}L. S. Kisslinger and Z. Li,
Phys. Rev. Lett. {\bf 74}, 2168 (1995). 
\bibitem{is} B. L. Ioffe and A. V. Smilga, Nucl. Phys. 
{\bf B232}, 109 (1984). 
\bibitem{eiskol} J. M. Eisenberg and D. S. Koltun, ``Theory of Meson 
Interactions with Nuclei''
(John Wiley and Sons, New York) (1979). 
\bibitem{lsk}L. S. Kisslinger, Phys. Rev. {\bf 98}, 761 (1955). 
\bibitem{cofugr}T. D. Cohen, R. J. Furnstahl, and  D. K. Griegel, 
Phys. Rev. Lett. {\bf 67}, 961 (1991);
Phys. Rev. {\bf C46}, 1507 (1992); \\ T. D. Cohen, R. J. Furnstahl, D.
K. Griegel, and X. Jin, Phys. Rev. {\bf C47}, 2882 (1993); \\ T. D.
Cohen, R. J. Furnstahl, D. K. Griegel, X. Jin, and M. Nielsen, Phys.
Rev. {\bf C49}, 464 (1994). \bibitem{gasser2} J. Gasser and H.
Leutwyler, Phys. Reports {\bf 87}, 77 (1982). 
\bibitem{henpas}E. M. Henley and J. Pasupathy, Nucl. Phys. 
{\bf A556}, 467 (1992).
\bibitem{tran} T. P. Cheng, Phys. Rev. {\bf D38}, 2869 (1988); \\ J.
Gasser, M. E. Sainio, and A. Svarc, Nucl. Phys. {\bf B307}, 779 (1988).
\bibitem{jk} M. B. Johnson and L. S. Kisslinger, Phys. Rev. {\bf C52},
1022 (1995). \bibitem{drukrysk}E. G. Drukarev and M. G. Ryskin, Nucl.
Phys. {\bf A578}, 333 (1994); Z. Phys. {\bf A356}, 457 (1997).
\bibitem{eletsky}V. L. Eletsky, Phys. Lett. {\bf B245}, 229 (1990).
\bibitem{koike}Y. Koike, Phys. Rev. {\bf D48}, 2313 (1993).
\bibitem{weinberg}S. Weinberg, Phys. Lett. {\bf 251}, 288 (1990);
Nucl.Phys. {\bf B363}, 1  (1991); Phys. Lett. {\bf B295}, 114 (1992).
\bibitem{pcac} T. P. Cheng and L. F. Li, ``Gauge Theory of Elementary
Particle Physics'' (Clarendon Press, Oxford) (1992) 
\bibitem{dist}H. De Vries, C. W. De Jager, and C. De Vries, At. 
Data Nucl. Data Tables {\bf 36}, 495 (1987); \\
M. B. Johnson and G. R. Satchler, Ann. Phys. (N.Y.) {\bf 248}, 
134 (1996).

\end{references}
\end{document}